# Photon Fluence and Dose Estimation in Computed Tomography using a Discrete Ordinates Boltzmann Solver

Edward T. Norris, and Xin Liu

*Abstract*—In this study, cone-beam single projection and axial CT scans are modeled with a software package – DOCTORS, which solves the linear Boltzmann equation using the discrete ordinates method. Phantoms include a uniform 35 cm diameter water cylinder and a non-uniform abdomen phantom. Series simulations were performed with different simulation parameters, including the number of quadrature angles, the order of Legendre polynomial expansions, and coarse and fine mesh grid. Monte Carlo simulations were also performed to benchmark DOCTORS simulations. A quantitative comparison was made between the simulation results obtained using DOCTORS and Monte Carlo methods. The deterministic simulation was in good agreement with the Monte Carlo simulation on dose estimation, with a root-mean-square-deviation (RMSD) difference of around 2.87%. It was found that the contribution of uncollided photon fluence directly from the source dominates the local absorbed dose in the diagnostic X-ray energy range. The uncollided photon fluence can be calculated accurately using a 'ray-tracing' algorithm. The accuracy of collided photon fluence estimation is largely affected by the pre-calculated multigroup cross-sections. The primary benefit of DOCTORS lies in its rapid computation speed. Using DOCTORS, parallel computing using GPU enables the cone-beam CT dose estimation nearly in real-time.

## I. INTRODUCTION

IN X-ray attenuation-based CT imaging, the mechanisms responsible for a material's attenuation are primarily the photoelectric effect and Compton scattering [1-3]. These interactions determine the energy transfer between photons and material as well as the photon distribution throughout the CT system. A complete description of the photon distribution and energy transfer is essential for estimating patient dose and for designing an optimized CT system. Stochastic methods (e.g., Monte Carlo simulation) have been used extensively in the past [4-20], and are generally considered to be the gold standard for estimating photon distributions and CT doses. However, they require a large number of particle histories and, therefore, a lengthy computation time is needed to reduce statistical uncertainty to an acceptable level. There is, however, no statistical error associated with deterministic methods, so they can be comparatively efficient in large regions where the highly resolved spatial fluence must be known to within a tight uncertainty bound. While hybrid stochastic-deterministic methods advantageously combine Monte Carlo and deterministic techniques and are more computationally efficient than a pure Monte Carlo simulation, they result in a cumbersome computational framework, due to the combination of two different methodologies, and can also have lengthy computation times [21-22].

We have explored three methodologies including Monte Carlo, hybrid Monte Carlo, and deterministic methods to solve photon transport problems [23-27], and have found that the deterministic method provides accurate results that are comparable to a Monte Carlo simulation. In addition, the deterministic method has the highest computational efficiency among the three methodologies. Although deterministic photon dose estimation has been widely used in the field of radiation therapy [28-32], its use has not been fully investigated for CT imaging. Because CT imaging is fundamentally different from radiation therapy, in terms of X-ray photon energy, interaction mechanisms, beam shape, and source trajectory, a CT-specific approach to deterministic photon dose calculation is required.

To date, we have developed several deterministic simulations of a CT system and its subcomponents [25-27]. In the course of these efforts, we discovered that the deterministic solution of the linear Boltzmann equation, based on the discrete ordinates method (i.e., $S_N$ method), is the most promising method, due to its scalability and parallelizability. Computer codes that are based on the discrete ordinates method have been extensively used in radiation shielding calculations and nuclear reactor analyses and, recently, have been utilized in clinical radiation therapy calculations. They have not, however, been applied to diagnostic imaging. Recently, we have developed a software application called DOCTORS (**D**iscrete **O**rdinate **C**omputed **TO**mography and **R**adiography **S**imulator) [33]. In this paper, we examined the accuracy and runtime of DOCTORS to compute energy-resolved photon fluence and dose distribution

This work was supported in part by the U.S. Nuclear Regulatory Commission under Grant NRC-HQ-13-G-38-0026.

Edward T. Norris was with Missouri University of Science and Technology, Rolla, MO 65401 USA. He is now with the Department of Energy, Washington DC (e-mail: etnc6d@ mst.edu).

Xin Liu is with Missouri University of Science and Technology, Rolla, MO 65401 USA. (e-mail: xinliu@mst.edu).

of a cone-beam CT with uniform and non-uniform phantoms.

## II. METHODS

The photon transport process can be described by the steady-state linear Boltzmann transport equation. The steady-state linear Boltzmann transport equation is given by the following: [34]

$$[\hat{\Omega} \cdot \vec{\nabla} + \sigma_t(\vec{r}, E)]\varphi(\vec{r}, E, \hat{\Omega}) = \iint_{4\pi} \sigma_s(\vec{r}, E' \to E, \hat{\Omega} \cdot \hat{\Omega}') \cdot \varphi(\vec{r}, E', \hat{\Omega}')d\hat{\Omega}'dE' + S(\vec{r}, E, \hat{\Omega}) \quad (1)$$

where, $\varphi(\vec{r}, E', \hat{\Omega}')$ is the angular fluence (photons/cm$^2$) at position $\vec{r}$, with energy $E$, and direction $\hat{\Omega}$; $\sigma_t(\vec{r}, E)$ is the total macroscopic interaction cross section, including scattering and absorption cross sections; $S(\vec{r}, E, \hat{\Omega})$ is the external source which simplifies the actual physical source; $\sigma_s(\vec{r}, E' \to E, \hat{\Omega} \cdot \hat{\Omega}')$ is the macroscopic differential scattering cross section which represents the probability that photons at position $\vec{r}$ are scattered from energy $E'$ and direction $\hat{\Omega}'$ to energy $E$ and direction $\hat{\Omega}$; $\hat{\Omega} \cdot \hat{\Omega}'$ is the cosine of the scattering angle. The left side of the Equation (1) represents photon loss in a differential volume at position $\vec{r}$, and the right side represents the gain of photons in the same differential volume and position. Thus, the linear Boltzmann transport equation actually describes the photon conservation of a fixed volume in a steady state.

### A. Discrete Ordinates Method

Analytic solutions of the Boltzmann transport equation can only be obtained for the simplest problems. Realistic, multidimensional, and energy-dependent problems must be solved numerically. The discrete ordinates method discretizes the continuous variables, $\vec{r}$, $E$, and $\hat{\Omega}$, onto a discrete phase space so that Equation (1) can be solved numerically [34]. Detailed derivation of discretized Boltzmann equation can be found in our previous publication [26]. Here, only the final discretized linear Boltzmann transport equation is given

$$[\hat{\Omega}_n \cdot \vec{\nabla} + \sigma_{i,j,k}^g]\phi_{i,j,k,n}^g = \sum_{n'=1}^{N} \sum_{g'=g}^{G-1} \sigma_{s,nn'}^{gg'} \phi_{i,j,k,n'}^{g'} \omega_n + S_{i,j,k,n}^g \quad (2)$$

where, $n$ represents the discrete direction; $i$, $j$, and $k$ represent the 3D spatial discrete mesh grid; $g$ represents the energy group; $\sigma_{s,nn'}^{gg'}$ is energy-group wised scattering cross section which is often expressed as a function of Legendre expansion [26]. The angular fluence $\phi_{i,j,k,n}^g$ in equation (2) can be solved numerically with proper boundary conditions. In our study, vacuum boundary conditions are applied. The scalar fluence at each voxel $(i, j, k)$ is obtained by integration over all of the discrete angles and approximated by the quadrature formula,

$$\phi_{i,j,k}^g = \sum_{n=1}^{N(N+2)} w_n \phi_{i,j,k,n}^g \quad (3)$$

where $N$ is the discrete ordinates order that are referred as $S_N$ quadratures which results in a total of $N(N+2)$ directions; $w_n$ is the weight associate with each discrete direction. Once the group photon fluence $\phi_{i,j,k}^g$ is solved, the absorbed dose or kerma can be obtained by applying the conversion factors from ICRP publications [35]. The collision kerma is usually considered equal to the absorbed dose due to the low X-ray energy used in CT scanning [14]. The accuracy of the discrete ordinates method depends on the number of discrete angles, energy groups, and the size of the mesh grid. However, a large number of discrete angles, energy groups, and a mesh grid would inevitably slow down the computation. Thus, a trade-off between accuracy and computation speed has to be evaluated when performing a simulation.

In addition, discrete ordinates methods suffer from 'ray-effect' in a weakly scattered media [34]. Ray-effect arises due to the restriction of particle transport to a set of discrete directions. To mitigate 'ray-effect', the first-collision source method was employed [36]. The fluence in each voxel is composed of uncollided photons directly from the source and collided photons scattered from other voxels, energies, and directions. Thus, equation (1) can be rewritten as two equations solving for the uncollided fluence $\phi_u$, and collided fluence $\phi_c$, independently. Both $\phi_u$, and $\phi_c$ obey the linear Boltzmann transport equation and the sum of $\phi_u$ and $\phi_c$ is the total fluence $\phi$. The uncollided photon transport equation has no scatter term since scattered particles are not considered in the uncollided fluence. The lack of a scatter term makes the uncollided fluence computable with a raytracing algorithm. The uncollided fluence is then used to compute the first-collision source. The first-collision source is used in the same way as the external source as shown in equations (1) and (2) to solve for the collided fluence in each voxel.

### B. The Discrete Ordinates Boltzmann Solver

We have recently developed dedicated discrete ordinates software, named DOCTORS, which computes photon fluence distribution and equivalent dose in a patient using the discrete ordinates method. The graphical user interface (GUI) of DOCTORS is shown in Figure 1. The first tab is the geometry input where a user provides a 3-D reconstructed image, such as DICOM (digital imaging and communications in medicine) images or raw CT numbers. As soon as the data is read in, it is converted into a series of dosimetrically equivalent materials representative of a human patient. Although there is no direct relation between CT numbers and tissue types, CT numbers can be converted to tissue types quite accurately based on a stoichiometric calibration [37-39].

The next three tabs identify the cross-section dataset, quadrature, anisotropy treatment, and whether GPU is used for the photon transport computation. The final tab, as shown in Figure 1, defines the X-ray source. A user can select a source type from a number of built-in options available for analysis, including point sources, fan beams, and cone beams. Multi-fan beams and multi-cone beams can be arranged about an object to mimic the source rotation in a CT scan. Each source type is described by its position and energy distribution. Two additional parameters, $\varphi$ and $\theta$, describe the azimuthal and polar angles subtended by the fan or cone beam, respectively.

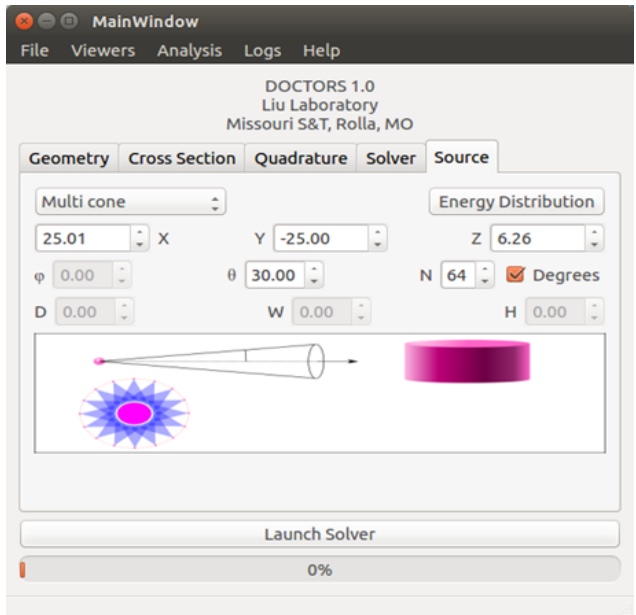

Fig. 1. The graphical user interface of DOCTORS.

Once all necessary data is loaded, the "Launch Solver" button becomes active and will remain so until the user creates a conflicting set of input that would prevent the solver from being able to run. When the user clicks on the "Launch Solver" button, the computation process begins and the photon fluence information will be displayed in the output dialog. Currently, DOCTORS cannot automatically identify specific organs, thus it is only able to compute the equivalent dose or whole-body effective dose via dose deposition. Organ-specific effective dose can be converted from the equivalent dose manually if specific organs could be identified.

*C. Verification against Monte Carlo Simulation*

The Monte Carlo simulation is considered the gold standard for patient dose estimation. DOCTORS has the capability to automatically generate an input file for the Monte Carlo simulation, using CT data and source specification provided by the user. The user can easily verify the results obtained from DOCTORS by comparison to a Monte Carlo simulation with the exact same input parameters. This feature greatly reduces the amount of work required to manually generate a Monte Carlo simulation input file. Currently, the only supported file format for a Monte Carlo simulation is the MCNP [40] input file format. MCNP, a widely-used Monte Carlo simulation package, has more than 50 years of history. The version of MCNP used in this study is MCNP6. Only analog Monte Carlo (i.e. no variance reduction techniques applied) simulation were performed in order to achieve a perfect analog to the photon transport in the patient. To achieve good statistics (i.e. <5% uncertainty), $1\times10^9$ histories were used in all Monte Carlo simulations.

The verification processes are divided into two categories: single projection verification and CT scan verification. The single projection verification compares the calculated collided and uncollided photon fluence in each energy group as well as the total effective dose for a single projection from a point source between DOCTORS and MCNP6 simulations. The CT scan verification compares the same quantities of DOCTORS and MCNP6 for 16 concurrent projections placed uniformly around the phantom, which mimic a CT scan [20]. All simulations were performed on a personal computer (PC) with an Intel i7-5960X processor and a base clock speed of 3.5 GHz.

Two source types are used in MCNP6 to benchmark DOCTORS. For the single projection verification, an isotropic point source was used. For the CT scan verification, 16 multi-cone beam sources were used. The cone angle is 30º for each cone beam. The source spectrum definition in MCNP6 was directly imported from DOCTORS' user input automatically. The cutoff energy of photon transport is 1 keV. Since the current version of DOCTORS does not simulate electron transport, electron transport was turned off in MCNP6 simulations. The cross-section library used in MCNP6 is the photon library from ENDF/B VII.0 [41]. All other transport parameters were set to the default values [40]. The 3D geometry and material type and density in each cell were directly imported from DOCTORS automatically. The tally type used is the photon fluence averaged over a cell (F4 tally). The photon fluence results are normalized to per source particle. Both collided and uncollided photon fluence were generated for each cell at specified energy bins.

*D. Deterministic Simulation Setup*

Two computational phantoms were used for verification of the DOCTORS code. The first phantom is a water cylinder phantom with 35 cm diameter and 12.5 cm in length. The water phantom is located at the center of a rectangular space with dimensions 50 x 50 x 12.5 cm in the x, y and z directions, respectively. The second phantom is about same dimension, but of a more realistic abdomen phantom. The data emulates a CT scan of a patient's midsection where the liver would be located.

The original phantom data for both the water phantom and the abdomen phantom are 256 x 256 x 64 meshes which contain 4,194,304 total voxels. Each voxel is a cube with sides of 1.95 mm. Though MCNP6 is capable of running such a mesh, the overhead of loading the mesh into memory and initiating the Monte Carlo solver can take on the order of hours, and it takes weeks to run the simulation. Therefore, a simplified geometry of 64 x 64 x 16 meshes which has only 65,536 voxels (the voxel size increased to 7.81 mm) was used to speed up Monte Carlo simulation. Theoretically, the mesh size should have little impact on the numerical solution of the discretized transport equation as long as the mesh size is much smaller than the mean-free-path of photons. However, large mesh size will inevitably degrade the resolution of photon fluence and dose map of the object. The mean-free-path of a photon at 20 keV is 13.9 mm in pure water, and monotonically increased to 60.6 mm at 100 keV..

In this study, four different quadrature sets $S_2$, $S_4$, $S_6$, and $S_8$ were used in DOCTORS, which represented 8, 24, 48, and 80 directions, respectively. The order of Legendre polynomial expansion varied from 0 to 1. The peak energy of the Bremsstrahlung radiation used in the simulations was 100 keV, and seven photon energy groups with energy boundaries 10, 20, 30, 45, 60, 70, 75, 100 keV were used. The multigroup cross sections used in DOCTORS are a part of the 47 gamma energy groups created by Oak Ridge National Laboratory (ORNL) from the ENDF/B-VII.0 cross section data library [41].

*E. GPU Acceleration*

Although the deterministic simulation is very efficient on a single CPU-based computer, a large number of discrete angles, energy groups, and voxels would inevitably slow down the computation. Thus, it is worth investigating parallel computing techniques that take advantage of popular computer architectures, such as the GPU. In the current version of DOCTORS, parallel ray-tracing and voxel sweeping algorithms were implemented on a single GPU architecture, using CUDA [42] language. The GPU card used here is NVidia GeForce GTX Titan-z which actually has two GPUs on one card. Since one of the GPUs is mainly used for display purposes, only one GPU is used for computation in the current implementation of DOCTORS. The performance of GPU acceleration was benchmarked for X-ray single projections and CT scans of the two phantoms with two different mesh grids (64 x 64 x 16 and 256 x 256 x 64).

## III. RESULTS

*A. Single Projection Verification with a Water Phantom*

Single projections from an isotropic point source on top of the phantom were simulated using both MCNP6 and DOCTORS. The uncollided and collided photon fluence distributions in each energy group of the central slice calculated by the two different methodologies were compared using RMSD (root mean square deviation). The RMSD results are listed in Table 1.

For collided photon fluence simulated by DOCTORS, eight different quadrature and Legendre parameter sets were tested. Since the uncollided photon fluence is calculated by a ray-tracing algorithm in DOCTORS which is independent of quadrature and Legendre parameters, only one set of uncollided photon fluence was calculated and compared to the MCNP6 simulations. In addition to the photon fluence comparison, total effective doses contributed from both collided and uncollided photons were also compared. The group-wise conversion factors that convert photon fluence to effective dose was obtained from ICRP116 [35] using linear interpolation. To simplify the calculation, the absorbed dose in air is neglected. Since the results of MCNP were normalized per source particle, the unit of the photon fluence and the total effective dose are number of photons per source particle and pSv per source particle, respectively.

The computer run time of MCNP6 was 645.33 minutes for a simulation with $1 \times 10^9$ histories. A large number of histories is necessary to achieve <5% statistical uncertainty in most of voxels. Nevertheless, the photon fluence simulated by MCNP6 in the energy group of 10 keV to 20 keV was discarded due to high statistical uncertainties. For DOCTORS, the computer run time varied based on the simulation parameters. A longer computation time was needed for the higher order quadrature sets and Legendre polynomials. The computer run time for different DOCTORS simulations are summarized in Table 2.

**Table 1.** Water phantom single projection photon fluence and total effective dose comparison

| Group (keV) | Collided Photon Fluence RMSD | | | | | | | | Uncollided RMSD |
|---|---|---|---|---|---|---|---|---|---|
| | $S_8P_0$ | $S_6P_0$ | $S_4P_0$ | $S_2P_0$ | $S_8P_1$ | $S_6P_1$ | $S_4P_1$ | $S_2P_1$ | |
| 75~100 | 37.29% | 36.70% | 36.05% | 33.07% | 35.87% | 35.62% | 34.89% | 37.79% | 3.70% |
| 70~75 | 32.91% | 32.37% | 31.49% | 30.16% | 36.86% | 36.27% | 35.81% | 37.56% | 3.18% |
| 60~70 | 27.39% | 26.65% | 26.83% | 35.97% | 28.85% | 27.44% | 27.81% | 39.71% | 3.02% |
| 45~60 | 8.56% | 8.81% | 10.56% | 27.76% | 5.39% | 5.66% | 7.16% | 28.09% | 3.32% |
| 30~45 | 18.83% | 18.59% | 19.66% | 35.89% | 18.60% | 18.19% | 19.34% | 36.28% | 3.56% |
| 20~30 | 25.36% | 25.28% | 25.22% | 27.11% | 25.13% | 24.86% | 24.88% | 26.97% | 7.59% |
| **Total Dose** | 5.02% | 4.84% | 4.89% | 7.06% | 6.05% | 5.72% | 5.69% | 8.19% | - |

* $P_n$ represents the n-th order of Legendre polynomial expansion.

**Table 2**. Single projection DOCTORS runtime for water and abdomen phantoms

| Quadrature | $S_2P_0$ | $S_4P_0$ | $S_6P_0$ | $S_8P_0$ | $S_2P_1$ | $S_4P_1$ | $S_6P_1$ | $S_8P_1$ |
|---|---|---|---|---|---|---|---|---|
| Water (minute) | 0.16 | 0.35 | 0.66 | 1.08 | 0.53 | 3.28 | 16.18 | 48.87 |
| Abdomen (minute) | 0.11 | 0.29 | 0.56 | 0.93 | 0.38 | 2.91 | 13.78 | 42.65 |

*B. Single Projection Verification with an Abdomen Phantom*

The same single projection verification process was performed on the abdomen phantom. The RMSD results are listed in Table 3. The computer runtime of MCNP6 simulation is 697.35 minutes with $1 \times 10^9$ histories. The computer run time of DOCTORS are summarized in Table 2.

**Table 3.** Abdomen phantom single projection photon fluence and total effective dose comparison

| Group (keV) | Collided Photon Fluence RMSD | | | | | | | | Uncollided RMSD |
|---|---|---|---|---|---|---|---|---|---|
| | $S_8P_0$ | $S_6P_0$ | $S_4P_0$ | $S_2P_0$ | $S_8P_1$ | $S_6P_1$ | $S_4P_1$ | $S_2P_1$ | |
| 75~100 | 30.39% | 30.09% | 29.70% | 24.75% | 24.87% | 24.71% | 25.01% | 28.30% | 3.08% |
| 70~75 | 26.92% | 26.36% | 25.40% | 23.19% | 31.01% | 30.67% | 30.40% | 31.04% | 3.29% |
| 60~70 | 50.24% | 51.43% | 53.02% | 65.19% | 48.07% | 49.21% | 50.01% | 61.55% | 3.43% |
| 45~60 | 8.73% | 9.01% | 10.72% | 23.39% | 7.74% | 7.90% | 9.69% | 23.79% | 3.59% |
| 30~45 | 9.51% | 9.83% | 12.52% | 28.34% | 9.56% | 9.84% | 12.50% | 28.60% | 4.45% |
| 20~30 | 17.38% | 16.73% | 17.26% | 28.31% | 16.76% | 16.19% | 16.82% | 28.50% | 4.91% |
| Total Dose | 3.96% | 3.88% | 4.51% | 10.21% | 4.00% | 3.81% | 4.18% | 10.22% | - |

### C. CT Scan Verification with Water Phantom

The CT scan simulation was realized by uniformly placing 16 cone beam sources around the water phantom. The cone angle was 30 degrees to cover the whole phantom. From the single projection verification process, it was found that using the $S_6$ quadrature and zeroth order Legendre polynomial expansion ($P_0$) generated the most accurate effective dose in DOCTORS. Thus, $S_6P_0$ was chosen to simulate a CT scan. The comparison between DOCTORS and MCNP6 was quantified by RMSD values of the central slice, which are summarized in Table 4. The computer runtime of the MCNP6 simulation was 1189.35 minutes with $1 \times 10^9$ histories whereas it only took 0.81 minutes for DOCTORS with the quadrature set $S_6P_0$. The MCNP6 relative uncertainty map and the relative difference map for collided photon fluence is shown in Figure 2. The relative difference is defined as the difference between DOCTORS and MCNP6 and normalized by MCNP6 values. The field-of-view (FOV) of the difference map is set to 35 cm. The MCNP6 relative uncertainty map and the relative difference map for uncollided photon fluence are shown in Figure 3. The total effective dose calculated by DOCTORS and MCNP6 as well as the difference map are shown in Figure 4.

**Table 4.** CT scan photon fluence and total effective dose comparison of a water phantom

| Group (keV) | RMSD | |
|---|---|---|
| | Collided $S_6P_0$ | Uncollided |
| 75~100 | 22.11% | 2.47% |
| 70~75 | 25.43% | 1.81% |
| 60~70 | 18.06% | 1.50% |
| 45~60 | 9.36% | 1.72% |
| 30~45 | 11.44% | 2.46% |
| 20~30 | 24.93% | 7.35% |
| Total Dose | 5.11% | |

### D. CT Scan Verification with Abdomen Phantom

The CT scan simulation was repeated on the abdomen phantom using the same parameters. The RMSD values are listed in Table 5. The computer runtime of MCNP6 simulation was 1207.46 minutes with $1 \times 10^9$ histories whereas it only took 0.78 minutes for DOCTORS with the quadrature set $S_6P_0$. The MCNP6 relative uncertainty map and the collided photon fluence relative difference map are shown in Figure 5. The MCNP6 relative uncertainty map and the uncollided photon fluence relative difference map are shown in Figure 6. The total effective doses calculated by MCNP6 and DOCTORS with quadrature set S6P0 as well as the relative difference map are shown in Figure 7.

**Table 5.** CT scan photon fluence and total effective dose comparison of an abdomen phantom

| Group (keV) | RMSD | |
|---|---|---|
| | Collided $S_6P_0$ | Uncollided |
| 75~100 | 28.14% | 1.60% |
| 70~75 | 23.68% | 1.23% |
| 60~70 | 17.59% | 1.17% |
| 45~60 | 6.52% | 1.36% |
| 30~45 | 8.68% | 2.29% |
| 20~30 | 16.19% | 6.76% |
| Total Dose | 2.87% | |

### E. GPU Acceleration Benchmark

Parallel ray-tracing and 3D voxel sweeping algorithms for X-ray single projections and CT scans were implemented on GPU architectures in DOCTORS. The GPU results are nearly identical to the results obtained from CPU. The maximum difference is between $\pm 0.02\%$. Table 6 summarizes the runtime required for the CPU-only version of DOCTORS and the GPU accelerated version. The quadrature set used in DOCTORS is $S_6P_0$. The speedup of the GPU over the CPU is also given in Table 6.

**Table 6.** GPU runtime benchmark on coarse (64x64x16) and fine mesh (256x256x64) grids

| Mesh Time (seconds) | Water Phantom | | | |
|---|---|---|---|---|
| | Coarse Mesh Single Projection | Coarse Mesh CT Scan | Fine Mesh Single Projection | Fine Mesh CT Scan |
| CPU | 39.5 | 48.8 | 5076.0 | 7393.9 |
| GPU | 2.8 | 3.2 | 66.8 | 364.4 |
| Speedup | 14.1 | 15.1 | 76.0 | 20.3 |

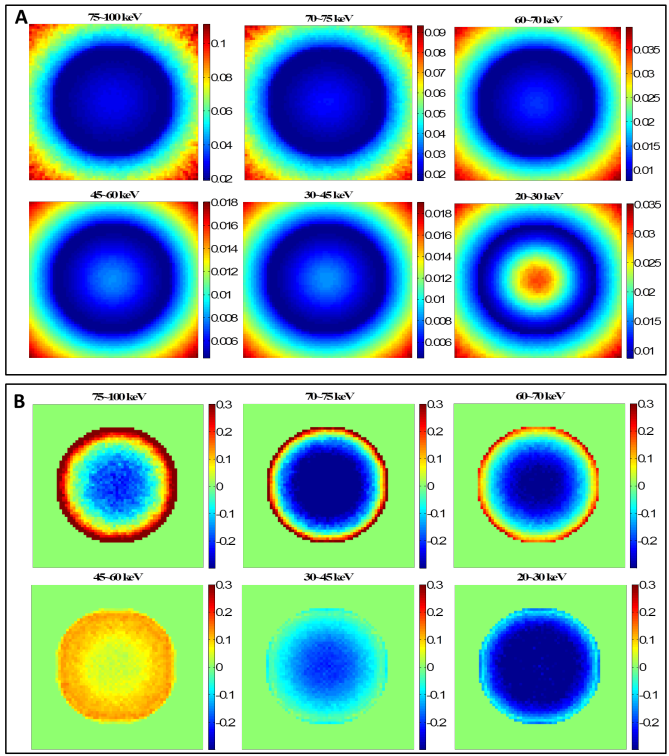

Fig. 2. Water phantom CT scan collided photon fluence simulations. A. MCNP6 collided photon fluence relative uncertainty map; B. Collided photon fluence relative difference map.

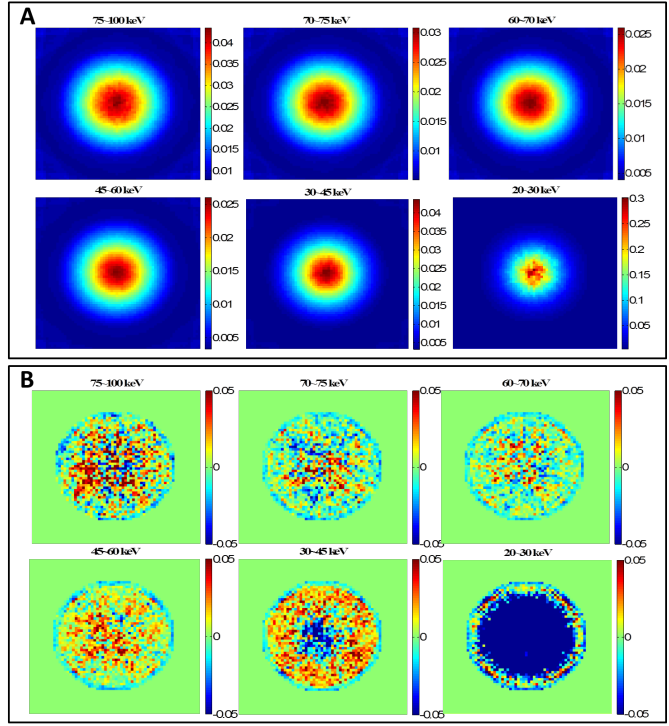

Fig. 3. Water phantom CT scan uncollided photon fluence simulations. A. MCNP6 uncollided photon fluence relative uncertainty map; B. Uncollided photon fluence relative difference map.

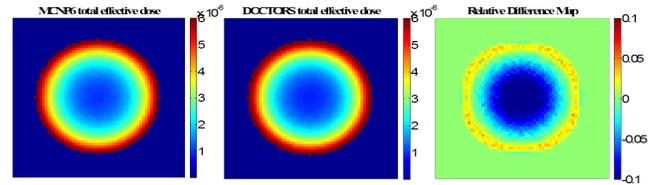

Fig. 4. Cross-section view of the total effective dose (pSv per source particle) from a CT scan in a water phantom calculated by MCNP6 and DOCTORS with quadrature set $S_6P_0$.

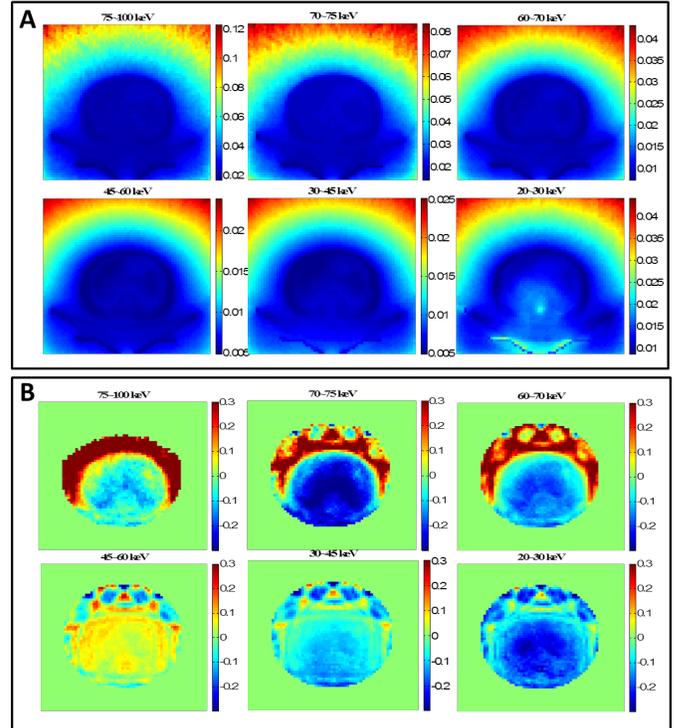

Fig. 5. Abdomen phantom CT scan collided photon fluence simulations. A. MCNP6 collided photon fluence relative uncertainty map; B. Collided photon fluence relative difference map.

## IV. DISCUSSION

As shown by the simulation results, both single projections and CT scans, the effective dose distribution calculated by DOCTORS was in good agreement with MCNP simulations. As shown in Tables 1, 3, 4, and 5, there are relative large discrepancies of DOCTORS simulations versus the MCNP simulation in collided photon fluence distributions for energy range 60~100 keV. However, the uncollided photon fluence directly from the source is dominant in the upper energy range (60~100 keV). The uncollided photon fluence was calculated analytically in DOCTORS such that it has very small discrepancies versus the MCNP simulations, as shown in the difference maps. Since the total effective dose in a voxel is composed of contributions from both collided and uncollided photon fluence, the discrepancies of effective dose between DOCTORS and MCNP simulations are much smaller.

As shown in Table 1, 3, 4, and 5, the discrepancies of DOCTORS simulations versus the MCNP simulation in the collided photon fluence distributions are consistent with certain energy groups across different quadrature sets, phantoms, and number of X-ray sources. In addition, the difference maps show that DOCTORS overestimates the collided photon fluence at the periphery and underestimates the collided photon fluence at the center of the object. This indicates that the large discrepancies are most likely caused by the multigroup cross sections. The multigroup cross sections used in DOCTORS were taken directly from ORNL's 47 gamma energy groups created from the ENDF/B VII data library. The 47-group photon cross section library was originally created for shielding analysis in nuclear engineering applications, which cover photon energies from 10 keV to 20 MeV. A similarity exists between gamma shielding analysis and photon transport in CT imaging since the most common shielding materials in nuclear engineering applications are water and concrete. However, multigroup cross sections are also affected by the object's geometry. ORNL's multigroup cross sections are only optimized for the nuclear engineering relevant problems. To develop an optimized multigroup cross section data library for clinical CT relevant problems is the focus of our future work.

To select an appropriate quadrature set, a series of simulations was performed. The results are summarized in Tables 1~4. It is found that low order quadrature set resulted in large errors, while high order quadrature set greatly increased computation time. The most suitable quadrature set is $S_6$ and zeroth order of Legendre expansion. The zeroth order Legendre expansion ($P_0$) indicates that isotropic scattering is dominant in most of the energy groups for photon energy range 20~100 keV. However, anisotropic scattering treatment may be required for photon energy greater than 100 keV.

As shown in Table 6, the computation time of DOCTORS was less than a minute for a single projection in a physical volume of 50cm x 50cm x 12.5cm on an ordinary personal computer. This computation time can be further decreased to less than 3 seconds by employing a parallel computing technique using GPU. For an axial CT scan, the computation time is still less than a minute on a personal computer using one CPU, and the computation time is decreased to about 3 seconds using GPU parallel computing..

One drawback of DOCTORS is the ray-effect artifact generated by the nature of discrete ordinates method. As shown in Figure 5, there are triangular shape artifacts in the air region around the phantom. These artifacts can also be seen in the difference map in Figure 5. Although the first-collision source method is very effective to remove ray-effect in the non-void region, it is not effective in the very weakly scattering air region outside the phantom. A simple method to remove the ray-effects artifact in the air region is to set the effective dose to be zero in air as shown in Figure 7.

Another drawback of the current version of DOCTORS is that specific organs cannot be automatically identified, thus it is only able to compute the equivalent dose or whole-body effective dose. If specific organs could be identified automatically, tissue specific weighting factors could be applied resulting in the effective dose to each organ which would be of greater clinical significance.

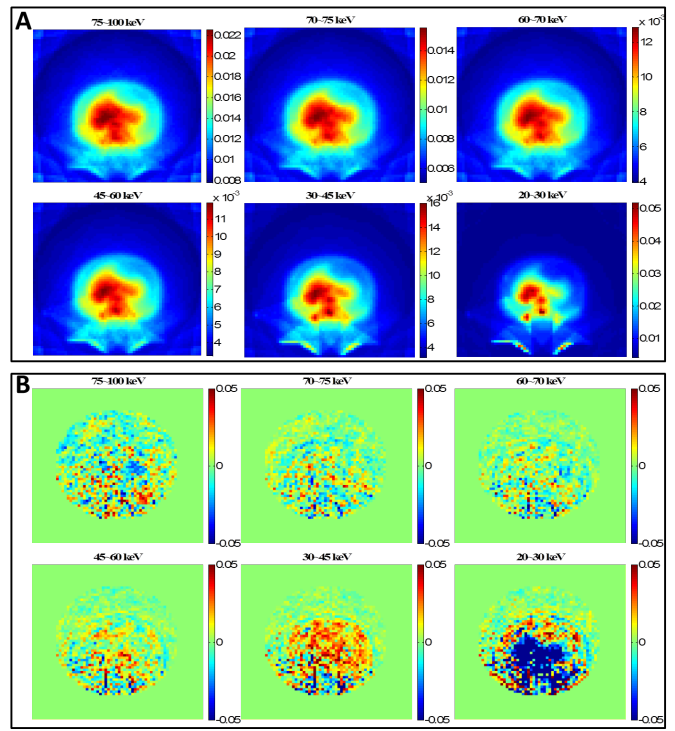

Fig. 6. Abdomen phantom CT scan uncollided photon fluence simulations. A. MCNP6 uncollided photon fluence relative uncertainty map; B. Uncollided photon fluence relative difference map.

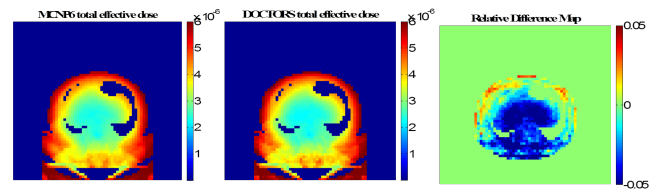

Fig. 7. Cross-section view of the total effective dose (pSv per source particle) from a CT scan in an abdomen phantom calculated by MCNP6 and DOCTORS with quadrature set $S_6P_0$.

V. CONCLUSION

Our simulation results showed that the discrete ordinates method can be used to estimate the photon fluence and dose distribution in uniform and non-uniform phantoms. The accuracy of the discrete ordinates method was close to that of a Monte Carlo simulation. The benefit of the discrete ordinates method is its computation efficiency. Energy-resolved photon fluence and dose distributions from a cone-beam axial scan can be calculated within a few seconds. A current limiting factor of our method is the multigroup photon cross section library which should be optimized for clinical CT imaging. Further optimization and utilization of this method in clinical setting are expected.